\documentclass[11pt]{article}
\usepackage[utf8]{inputenc} 
\usepackage{braket}
\usepackage{amsmath,latexsym}
\usepackage{dsfont}
\usepackage{physics}
\usepackage{indentfirst}
\usepackage{hhline}
\usepackage{amsfonts}
\usepackage[table]{xcolor}
\usepackage[english]{babel}
\usepackage{graphicx}
\usepackage{pgfplots}
\usepackage{authblk}
\usepackage{hyperref}
\usepackage{todonotes}
\usepackage{array}
\usepackage{hyperref}
\hypersetup{
    colorlinks=true,
    linkcolor=blue,
    filecolor=magenta,      
    urlcolor=cyan,
}
\pgfplotsset{width=9cm,compat=1.9}
\graphicspath{ {images/} }

\definecolor{cornellred}{rgb}{0.7, 0.11, 0.11}
\definecolor{cadmiumgreen}{rgb}{0.0, 0.42, 0.24}
\definecolor{darkcerulean}{rgb}{0.03, 0.27, 0.49}

\usepackage{stackengine}

\newcommand{\be}{\begin{equation}}
\newcommand{\ee}{\end{equation}}
\newcommand{\ben}{\begin{enumerate}}
\newcommand{\een}{\end{enumerate}}
\def\beqn{\begin{eqnarray}}
\def\eeqn{\end{eqnarray}}

\begin{document}

\vspace{5mm}

\centerline{\Large \bf On resummation of the irregular conformal block}
\vspace{5mm}

\centerline{S. O. Alekseev$^{2,3}$ and M. V. Litvinov$^{1,2,3}$}

\begin{center}
{\it ${}^{1}$ Skolkovo Institute of Science and Technology, 121205, Moscow, Russia}
\end{center}
\begin{center}
{\it ${}^{2}$ Moscow Institute of Physics and Technology, 141700, Dolgoprudny, Russia}
\end{center}
\begin{center}
{\it ${}^{3}$ Institute for Theoretical and Experimental Physics, 117218, Moscow, Russia}
\end{center}


\begin{abstract}
We discuss the resummation procedure of the superpotential in $4\dd$ $\mathcal{N}=2$ SYM theory without matter from the point of view of $2\dd$ Liouville conformal field theory, utilizing the AGT correspondence. We identify contributions of different descendants of the intermediate state, defining a conformal block in CFT, to the superpotential and answer the question, which descendants are responsible for the appearance of the branch cuts in the superpotential.
\end{abstract}

\section{Introduction}

The solution of Seiberg and Witten \cite{SW} of $\mathcal{N} = 2$ gauge theory used the constraints of special geometry of the moduli space of vacua to achieve understanding of the strong coupling dynamics of gauge theory. Direct evaluation of the Seiberg-Witten prepotential was accomplished by Nekrasov \cite{Nekrasov} by considering the theory in the $\Omega$-background $\mathbb{C}^2_{\epsilon_1,\epsilon_2}$, which provides the IR regularization by localizing the integrals over the instanton moduli spaces on a set of isolated fixed points. The Seiberg-Witten prepotential defining the low-energy effective action was obtained as the limit of the deformed partition function: $\mathcal{F}(\vec a,\vec \mu,\Lambda)=\lim\limits_{\epsilon_1,\epsilon_2 \to 0}\epsilon_1\epsilon_2 \log \mathcal{Z}(\vec a,\vec \mu,\Lambda;\epsilon_1,\epsilon_2)$. 

In \cite{NS} it was argued that the effective two dimensional $\mathcal{N}=(2,2)$ theory, described by the effective twisted superpotential $\mathcal{W}(\vec a,\vec \mu,\Lambda;\epsilon_1)=\lim\limits_{\epsilon_2\to 0}\epsilon_2 \log\mathcal{Z}(\vec a,\vec \mu,\Lambda;\epsilon_1,\epsilon_2)$, is related to some quantized algebraic integrable system, and the parameter $\epsilon_1=\hbar$ plays the role of the quantization parameter. For pure $U(2)$ gauge theory finding of the expectation values of the chiral trace operator is equivalent to finding of the spectrum of the Mathieu equation. However, particular periodicity properties of the chiral trace operator, following from the Mathieu equation, are not manifest when the superpotential is represented as the expansion in the instanton number. This problem was solved by an approriate procedure of trans-series resummation \cite{GorskyMS}, concerning reordering of the summation of trans-series near the poles. As a consequence of that procedure, the naive pole singularities in the superpotential transform into the cuts. There were several previous studies addressing the issue of singularities in the effective twisted superpotential. In \cite{Beccaria} the resummation procedure was performed for finite-gap $\mathcal{N} = 2^*$ $U(2)$ gauge theory.

According to the AGT correspondence \cite{AGT,2GaiDu}, the
Nekrasov partition function is identified with the particular conformal block in the Liouville theory, whose type depends on the matter content in the gauge theory. In particular, the scalar field expectation value corresponds to the dimension of the intermediate state in the conformal block and the poles in the superpotential correspond to the particular values of intermediate dimensions. Thus, a natural question arises: what is the meaning of the resummation procedure of the superpotential in the pure $U(2)$ gauge theory \cite{GorskyMS} from the point of view of the two-dimensional conformal field theory? The partition function for the pure gauge theory in $4\dd$ is equal to the so called irregular conformal block in two-dimensional theory, which is a norm squared of a certain coherent state, called Gaiotto state \cite{GaiottoIrreg}. The irregular conformal block for pure $SU(2)$ can be also obtained by a limiting procedure from the conformal block on the sphere with four insertions \cite{MMM}. The Mathieu equation on the Liouville theory side follows from the null vector decoupling equation (see e.g. \cite{Jap,Piatek}), which can be obtained by inserting a degenerate operator between two Gaiotto states. 
 
The paper is organized as follows. In Section 2 we reproduce known results concerning the procedure of obtaining the irregular conformal block from the 4-point conformal block in an appropriate limit and we also take the same limit in the Zamolodchikov recursion relation. In Section 3 we discuss the resummation procedure of the superpotential from the point of view of CFT and answer the question, which intermediate states contributing to the conformal block are responsible for the branch cuts in the superpotential. Some open questions concerning the interpretation of the resummation procedure in terms of AdS${}_3$/CFT${}_2$ correspondence are discussed in Conclusion.
\section{Preliminaries}
\subsection{Conformal block in the limit of large external dimensions}
\noindent According to the AGT correspondence \cite{AGT}, the instanton part of the Nekrasov partition function in $\mathcal N=2$ SYM with the gauge group $SU(2)$ and four matter multiplets in fundamental representation is equal to the 4-point conformal block in the Liouville conformal field theory on the Riemann sphere
\begin{equation}
\label{eq:AGT}
\mathcal{Z}_{\ \operatorname{inst}}^{SU(2)}(a,\vec \mu, \Lambda;\epsilon_1,\epsilon_2)=\mathcal{F}(c,\Delta,\Delta_i;z)    
\end{equation}
after the following identification of parameters between these two theories:
\renewcommand*{\arraystretch}{2}
\begin{center}
  \begin{tabular}{ | c | c | }
    \hline
    \textbf{4d gauge theory} & \textbf{2d Liouville theory}  \\ \hline \hline
    Deformation parameters $\epsilon_1, \epsilon_2$ & Central charge $\displaystyle b^2=\frac{\epsilon_2}{\epsilon_1}$  \\[1ex] \hline
    Vacuum expectation value $a$ & Internal dimension
    
    $\displaystyle\alpha=\frac{Q}{2}+\frac{a}{\sqrt{\epsilon_1\epsilon_2}}$  \\[1.5ex]
    \hline
     & External dimensions\\
    
    Masses of matter multiplets $\mu_i$ & $\displaystyle\alpha_1=\frac{Q}{2}+\frac{\mu_1-\mu_2}{2\sqrt{\epsilon_1\epsilon_2}},\,  \alpha_2=\frac{\mu_1+\mu_2}{2\sqrt{\epsilon_1\epsilon_2}},$\\
    
    & $\displaystyle\alpha_3=\frac{\mu_3+\mu_4}{2\sqrt{\epsilon_1\epsilon_2}},\, \alpha_4=\frac{Q}{2}+\frac{\mu_3-\mu_4}{2\sqrt{\epsilon_1\epsilon_2}}$  \\[1.5ex] \hline
    Dynamically generated scale  $\Lambda$ & Cross ratio $\displaystyle z= \frac{\Lambda^4}{\prod \mu_i}$  \\[1.5ex] \hline
  \end{tabular}
\end{center}
As usual, the standard parametrization of quantities is used in the Liouville theory:
\begin{equation*}
\begin{split}
&c=1+6Q^2,\ Q=b+b^{-1},\\
&\Delta=\alpha(Q-\alpha).
\end{split}
\end{equation*}
The resummation procedure in \cite{GorskyMS} was performed in the pure gauge theory, so we need to get rid off the matter multiplets on the l.h.s of the equation (\ref{eq:AGT}). It is known that on the r.h.s. this corresponds to considering the irregular conformal block, which can be obtained from the 4-point conformal block on the sphere in the limit of large conformal dimensions \cite{MMM}:
\begin{equation}
\label{eq:lim}
\begin{split}
&\Delta_2,\Delta_3 \to -\infty,\\
&z \to 0,\\
& \Delta_2\Delta_3\, z\ \text{is kept fixed}.
\end{split}
\end{equation}
One possible way to satisfy the first of these conditions is to put $\mu_i=\mu,\ i=1,..,4$ and to take the limit of large masses $\mu \to \infty$. Let us reproduce the conformal block in the limit of large conformal dimensions to establish notations. The identity operator in the conformal field theory containing primary fields labelled by their conformal dimensions $\Delta$ has the following form:
\be
\mathds{1}=\sum_{\Delta;\,Y',Y}L_{-Y'}\ket{\Delta}\big(\mathcal N^{-1}_{\,\Delta}\big)_{Y'Y}\bra{\Delta}L_{Y},
\ee
where the matrix $\mathcal N$ is the Gram matrix of all basis vectors in the Verma module of the vector $\ket{\Delta}$:
\begin{equation*}
(\mathcal{N}_\Delta)_{\,Y'Y}=\bra{\Delta}L_{Y'}L_{-Y}\ket{\Delta}.
\end{equation*}
Then plugging the identity operator into the 4-point correlation function allows one to pick out contributions of each conformal family to it (conformal blocks):
\begin{equation*}
\begin{split}
&\Braket{O_4(\infty)O_3(1)\,\mathds{1}\,O_2(z)O_1(0)}=\sum_{\Delta;\, Y',Y}\bra{0}O_4(\infty)O_3(1)L_{-Y'}\ket{\Delta} \big(\mathcal N^{-1}_{\,\Delta}\big)_{Y'Y} \bra{\Delta}L_{Y}O_2(z)O_1(0)\ket{0}=\\
&=\sum_\Delta C_{43\Delta} C_{\Delta21} F (c,\Delta,\Delta_i;z),
\end{split}
\end{equation*}
where 
\be
\label{eq:conf_block}
F(c,\Delta,\Delta_i;z)=\sum_{Y',Y} \frac{\bra{0}O_4(\infty)O_3(1)L_{-Y'}\ket{\Delta}}{C_{43\Delta}}\, \big(\mathcal N^{-1}_{\,\Delta}\big)_{Y'Y}\,\frac{\bra{\Delta}L_{Y}O_2(z)O_1(0)\ket{0}}{C_{\Delta21}}.
\ee
Because of the form of the correlator $\bra{\Delta}O_2(z)O_1(0)\ket{0}$, the conformal block $F(c,\Delta,\Delta_i;z)$ has the following expansion in powers of $z$:
\begin{equation*}
F(c,\Delta,\Delta_i;z)\equiv z^{\Delta-\Delta_1-\Delta_2} \mathcal F(c,\Delta,\Delta_i;z)= z^{\Delta-\Delta_1-\Delta_2} \sum_{k\ge 0} \mathcal F_k z^k.
\end{equation*}
To take the limit of large conformal dimensions (\ref{eq:lim}) in the expression (\ref{eq:conf_block}) for the conformal block, it is enough to calculate the 3-point correlators only, as the Gram matrix does not depend on external dimensions $\Delta_i$. The both 3-point correlators can be calculated explicitly with the use of the Virasoro algebra. For instance,
\be
\bra{\Delta}L_{m_1}...\,L_{m_n}O_2(z)O_1(0)\ket{0}=\mathcal L_{m_n}...\,\mathcal L_{m_1}\bra{\Delta}O_2(z)O_1(0)\ket{0},
\ee
where $\mathcal L_m= \Delta_2(1+m)z^m+z^{m+1}\partial_z$, so that:
\be
\label{eq:3ptcorr_1}
\begin{split}
&\bra{\Delta}L_{m_1}...\,L_{m_n}O_2(z)O_1(0)\ket{0}=z^{\sum_{i=1}^n m_i}\prod\limits_{i=1}^n\bigg[\Delta+\sum\limits_{j=1}^{i-1}m_j+m_i\Delta_2-\Delta_1\bigg]\bra{\Delta}O_2(z)O_1(0)\ket{0}.
\end{split}
\ee
The same procedure is applicable to the second 3-pt correlator in the formula (\ref{eq:conf_block}):
\begin{equation*}
\bra{0}O_4(\infty)O_3(1) L_{-m_n}...\, L_{-m_1}\ket{\Delta}=\mathcal L_{-m_n}...\,\mathcal L_{-m_1}\bra{0}O_4(\infty)O_3(z)\ket{\Delta}\big|_{z=1},
\end{equation*}
where now $\mathcal L_{-m}=-\Delta_3(1-m)z^{-m}-z^{-m+1}\partial_z$. So,
\begin{equation}
\label{eq:3ptcorr_2}
\begin{split}
&\bra{0}O_4(\infty)O_3(1) L_{-m_n}...\, L_{-m_1}\ket{\Delta}=\prod\limits_{i=1}^n\bigg[\Delta+\sum\limits_{j=1}^{i-1}m_j+m_i\Delta_3-\Delta_4\bigg]\bra{0}O_4(\infty)O_3(1)\ket{\Delta}.
\end{split}
\end{equation}
Now we combine the formulae (\ref{eq:conf_block}), (\ref{eq:3ptcorr_1}), (\ref{eq:3ptcorr_2}) together and take the limit of large conformal dimensions $\Delta_2,\Delta_3\to -\infty$ in the conformal block. As one can see from the expressions (\ref{eq:3ptcorr_1}), (\ref{eq:3ptcorr_2}), the more Virasoro operators are needed to create an intermediate state at a given level of the Verma module, the greater power of $\Delta_2$ or $\Delta_3$ is produced in the correlators. Therefore, of all the contributions to the conformal block that come from the level $N$, only that of the state $L^{N}_{-1}\ket{\Delta}$ survives in the limit under consideration:
\be
\label{eq:exp_fromOPE}
\mathcal F(c,\Delta;z)=\lim_{\substack{\mu\to \infty \\ z\to 0}} \mathcal F(c,\Delta,\Delta_i;z)= \sum_N \frac{\mu^{4N}}{(\epsilon_1\epsilon_2)^{2N}}\, \big(\mathcal N^{-1}_{\Delta}\big)_{\{1^N;1^N\}}\, z^N= \sum_N \frac{1}{(\epsilon_1\epsilon_2)^{2N}}\, \big(\mathcal N^{-1}_{\Delta}\big)_{\{1^N;1^N\}}\, \tilde z^N,
\ee
where in the last equality we introduced a new rescaled variable $\tilde z= \mu^4 z$, which is kept finite in accordance with (\ref{eq:lim}). Each term in the formula above has a manifest meaning in terms of OPE, namely that the $N$th order in $\tilde z$ comes from the OPE channel $O_2O_1 \to L_{-1}^N O_{\Delta}$. However, this formula is still not convenient for calculations as it involves the inverse Gram matrix. Although we need only one element of it at each level, to find this element one needs to know the whole Gram matrix at the level.

\subsection{Zamolodchikov's recursion relation}
\noindent For actual computations of conformal block coefficients, it is more convenient to use Zamolodchikov's recursion relation, in which we also take the limit of large conformal dimensions. According to Zamolodchikov \cite{Z1,Z2}, a 4-point conformal block on the sphere can be represented in the following form (see also \cite{Perlmutter}):
\begin{equation}
\label{eq:2nd_Z}
\mathcal{F}(c,\Delta,\Delta_i;z)=\bigg(\frac{16q}{z}\bigg)^{\Delta-\frac{c-1}{24}}(1-z)^{\frac{c-1}{24}-\Delta_2-\Delta_3}\,\theta_3(q)^{\frac{c-1}{2}-\sum \Delta_i}\ H(c,\Delta,\Delta_i;q),
\end{equation}
where the function $H(c,\Delta,\Delta_i,q)$ obeys the recurrence relation:
\begin{equation}
\label{eq:recur}
H(c,\Delta,\Delta_i;q)=1+\sum_{m,n\ge1}^\infty \frac{(16q)^{mn}R_{mn}(c,\Delta_i)}{\Delta-\Delta_{mn}(c)}\ H(c,\Delta_{mn}+mn,\Delta_i;q),
\end{equation}
$q(z)$ is an elliptic variable and $\theta_3(q)$ is the Jacobi theta function. In \cite{Z1} an explicit expression for the coefficients $R_{mn}(c,\Delta_i)$ was established:
\begin{equation}
\label{eq:rmn}
R_{mn}(c,\Delta_i)=-\frac{1}{2}\prod_{a,b}\frac{1}{l_{ab}}\ \prod_{j,k}\bigg(l_2+l_1-\frac{l_{jk}}{2}\bigg)\bigg(l_2-l_1-\frac{l_{jk}}{2}\bigg)\bigg(l_3+l_4-\frac{l_{jk}}{2}\bigg)\bigg(l_3-l_4-\frac{l_{jk}}{2}\bigg),
\end{equation}
The products run over the following sets of integers:
$a=-m+1,-m+2,\ldots,m;\ b=-n+1,-n+2,\ldots,n$ except for the pairs $(a,b)=(0,0)$ and $(m,n)$; $j=-m+1,-m+3,\ldots,m-3,m-1;\ k=-n+1,-n+3,\ldots,n-3,n-1$. In the last expression parameters $l_i$ are defined from conformal dimensions of external operators:
\begin{equation*}
\Delta_i=\frac{c-1}{24}+l_i^2
\end{equation*}
and $l_{jk}=j\alpha_++k\alpha_-$, where $\alpha_+$ and $\alpha_-$ are the parameters entering the expression for conformal dimensions of degenerate operators:
\begin{equation*}
\Delta_{mn}=\frac{c-1}{24}+\frac{1}{4}(m\alpha_++n\alpha_-)^2.
\end{equation*}
If one expands the function $H(c,\Delta,\Delta_i,q)$ in powers of $16q$ 
\begin{equation*}
H(c,\Delta,\Delta_i;q)=\sum_{K=0}^\infty H_K(c,\Delta,\Delta_i) (16q)^K,
\end{equation*}
then the recursion relation on the coefficients $H_K$, following from the recurrence relation (\ref{eq:recur}), is as follows:
\begin{equation*}
H_K(c,\Delta,\Delta_i)=\sum_{mn+N=K} \frac{ R_{mn}(c,\Delta_i)}{\Delta-\Delta_{mn}}\ H_N(c,\Delta_{mn}+mn,\Delta_i),\ H_0=1.
\end{equation*}
To take the limit of large conformal dimensions (\ref{eq:lim}) in the equation (\ref{eq:2nd_Z}), we first note that the prefactor in front of the function $H(c,\Delta,\Delta;q)$ on the r.h.s, tends to $1$ in this limit. To demonstrate how the limit can be taken in the function $H(c,\Delta,\Delta_i;q)$, let us write out the first few expansion coefficients $H_K$, leaving only leading powers of masses $\mu$ in them:
\begin{equation*}
\begin{split}
&H_0=1\\
&H_1=\frac{ R_{11}(c,\Delta_i)}{\Delta-\Delta_{11}}= \frac{\mu^4}{(\epsilon_1\epsilon_2)^2}\ \frac{ A_{11}(c)}{\Delta-\Delta_{11}}+...\\
&H_2=\frac{ R_{12}(c,\Delta_i)}{\Delta-\Delta_{12}}+\frac{ R_{21}(c,\Delta_i)}{\Delta-\Delta_{21}}+\frac{ R_{11}(c,\Delta_i)}{\Delta-\Delta_{11}}\cdot  R_{11}(c,\Delta_i)=\\ &\hspace*{220pt}=\frac{\mu^8}{(\epsilon_1\epsilon_2)^4} \bigg[\frac{ A_{12}(c)}{\Delta-\Delta_{12}}+\frac{ A_{21}(c)}{\Delta-\Delta_{21}}+\frac{\big[ A_{11}(c)\big]^2}{\Delta-\Delta_{11}}\bigg]+...
\end{split}
\end{equation*}
with the coefficients $A_{mn}(c)$ being equal to (cf. (\ref{eq:rmn}))
\begin{equation*}
    A_{mn}(c)=-\frac{1}{2}\prod_{a,b}\frac{1}{l_{ab}}.
\end{equation*}
Using the fact that $16q=z+O(z^2)$, defining the same rescaled coordinate $\tilde z=\mu^4 z$ as above and also the rescaled coefficients $\tilde H_K(c,\Delta)=(\epsilon_1\epsilon_2)^{2K}\lim_{\mu\to \infty}\mu^{-4K} H_K(c,\Delta,\Delta_i)$, one finds the following expression for the conformal block in the limit of large conformal dimensions:
\begin{equation}
\label{eq:recurr_rel_largem}
\begin{split}
&\mathcal{F}(c,\Delta;\tilde z)=\tilde H(c,\Delta,\tilde z)\equiv\lim_{\substack{\mu\to \infty \\ z\to 0}} H(c,\Delta,\Delta_i;q)= \sum_{K=0}^\infty \frac{1}{(\epsilon_1\epsilon_2)^{2K}}\, \tilde H_K(c,\Delta)\ \tilde z^K,
\end{split}
\end{equation}
where the coefficients $\displaystyle \tilde H_K(c,\Delta)$ obey now a simpler recurrence relation compared to (\ref{eq:2nd_Z}):
\begin{equation}
\label{eq:recurs_H}
\tilde H_K(c,\Delta)=\sum_{mn+N=K}\frac{ A_{mn}(c)}{\Delta-\Delta_{mn}}\ \tilde H_N(c,\Delta_{mn}+mn),\ \tilde H_0=1.
\end{equation}
Comparing two expressions for the conformal block in the limit of large conformal dimensions (\ref{eq:exp_fromOPE}) and (\ref{eq:recurr_rel_largem}), we obtain that
\begin{equation}
\big(\mathcal N^{-1}\big)_{\{1^N;1^N\}}=\tilde H_N(c,\Delta),
\end{equation}
and thus the element $\{1^N;1^N\}$ of the Gram matrix at the $N$th level of the Verma module can be calculated with the use of the recursion relation (\ref{eq:recurs_H}), which is simpler than inverting the Gram matrix. In the text below we write $H_K$ instead of $\tilde H_K$, omitting the tilde.
\section{Analysis of the resummation procedure}
\noindent The conformal block in the limit of large external dimensions reproduces the instanton part of the Nekrasov partition in the pure gauge theory
\begin{equation}
\label{eq:AGT2}
\mathcal{Z}_{\ \operatorname{inst}}^{SU(2)}(a, \Lambda;\epsilon_1,\epsilon_2)=\mathcal{F}(c,\Delta;z),
\end{equation}
and its logarithm, in particular, is equal to
\begin{equation}
\label{eq:thus}
\begin{split}
&\log \mathcal F(c,\Delta; z)=-\frac{\epsilon_1}{\epsilon_2}\bigg[\frac{2}{\nu^2-1}\ \frac{ z}{\epsilon_1^4}+\frac{1}{\epsilon_2}\cdot\frac{5\nu^2+7}{(\nu^2-1)^3(\nu^2-4)}\ \frac{ z^2}{\epsilon_1^8} +...\bigg]+O(\epsilon_2^{\, 0})=\\
&=-\frac{1}{\epsilon_2}\mathcal{W}(z)+O(\epsilon_2^{\, 0}),
\end{split}
\end{equation}
where $\mathcal{W}(z)$ is the superpotential mentioned in the introduction. First of all, we would like to know which of the intermediate states $L_{-1}^k \ket{\Delta}$ contribute to the $N$th order of the superpotential $\mathcal W(z)=\sum_{N} \mathcal{W}_N\,z^N$. To see this, we expand the coefficients $H_N$ in powers of $\epsilon$ (in what follows we put $\epsilon_1=1,\, \epsilon_2=\epsilon$). It follows from the formula (\ref{eq:recurr_rel_largem}) that the expansion of the coefficient $H_N$ in powers of $\epsilon$ starts with $\epsilon^{-N}$:
\begin{equation*}
H_N=\sum_{k=-N}^\infty a_{\, k}^{(N)} \epsilon^k.
\end{equation*}
For instance, the first three coefficients $H_K$ are expanded in powers of $\epsilon$ as follows:
\begin{equation}
\label{eq:coeff_exp_eps}
\begin{split}
&H_1=-\frac{2}{\epsilon(\nu^2-1)}+O\left(\epsilon^0\right),\\
&H_2=\frac{2}{\epsilon^2(\nu^2-1)^2}+\frac{3 (\nu^2-13)}{\epsilon(\nu^2-4)(\nu^2-1)^3 }+O\left(\epsilon^0\right),\\
&H_3=-\frac{4}{3\epsilon^3(\nu^2-1)^3}+\frac{2 (\nu^2+23)}{\epsilon^2(\nu^2-4) (\nu^2-1)^4}-\frac{4 \left(3 \nu^8-123 \nu^6+1459 \nu^4-1757
\nu^2-6782\right)}{3\epsilon(\nu^2-9) (\nu^2-4)^2 (\nu^2-1)^5}+O\left(\epsilon^0\right).
\end{split}    
\end{equation}
Consequently, taking into account that the most singular power of $\epsilon$, entering the logarithm of the conformal block, is $\epsilon^{-1}$, we have:
\begin{equation}
\label{eq:meaning_of_prep}
\begin{split}
&\log \mathcal F(z)=\frac{1}{\epsilon}\mathcal W(z)+O(\epsilon^0)=\bigg(\frac{a_{-1}^{(1)}}{\epsilon}+a_{\, 0}^{(1)}+...\bigg)z+\bigg(\frac{a_{-1}^{(2)}-a_{-1}^{(1)}a_{\, 0}^{(1)}}{\epsilon}+...\bigg)z^2+\\
&+\bigg(\frac{a_{-1}^{(3)}-\big[a_{-1}^{(1)}a_{\, 0}^{(2)}+a_{\, 0}^{(1)}a_{-1}^{(2)}+a_{1}^{(1)}a_{-2}^{(2)}\big]+\big[\big(a_{-1}^{(1)}\big)^2 a_{\, 1}^{(1)}+a_{-1}^{(1)} \big(a_{\, 0}^{(1)}\big)^2\big]}{\epsilon}+...\bigg)z^3+...
\end{split}
\end{equation}
So, we see that the $N$th order of the superpotential, expanded in powers of $z$, contains contributions from all the intermediate states $L_{-1}^k \ket{\Delta}$ with $k\le N$. Obviously, the $N$th order is a sum of all the terms $a_{\, p_1}^{(q_1)}a_{\, p_2}^{(q_2)}...\, a_{\, p_l}^{(q_l)}$ with $\sum p_i=-1, \sum q_i= N$. In what follows we omit  the parentheses and write $a_k^n$ instead of $a_k^{(n)}$.\\

Let us also compare poles in the conformal block $\mathcal F(z)$ and in the superpotential $\mathcal W(z)$. We have seen already that obtaining the superpotential requires expanding the coefficients $H_N$ in powers of $\epsilon$. What happens with poles in the intermediate dimension $\Delta$ after this procedure? All the poles which enter the coefficients $H_N$ are simple and are of the form $(\Delta-\Delta_{mn})^{-1}$, so they are labelled by two integers $m, n$. Expanding them in powers of $\epsilon$, one obtains:
\begin{equation*}
\frac{1}{\Delta-\Delta_{mn}}=\frac{4\epsilon}{(m+n\epsilon)^2-\nu^2}=4\epsilon \bigg[\frac{1}{m-\nu}-\frac{n\epsilon}{(m-\nu)^2}+...\bigg]\bigg[\frac{1}{m+\nu}-\frac{n\epsilon}{(m+\nu)^2}+...\bigg].
\end{equation*}
However, one sees that the poles of the same expression in the variable $\nu$ are labelled by one integer number $m$, but the poles of all orders in $\nu$ appear because of the expansion in $\epsilon$. From the point of view of the resummation made in \cite{GorskyMS} the most important terms in the superpotential are those which are the most singular in $\nu$ in each order $\mathcal{W}_N$ as these terms give logarithms and thus branch cuts after the resummation. The following question arises: which intermediate states $L_{-1}^k\ket{\Delta}$ contribute to these most singular terms responsible for branch cuts in the superpotential? To answer this question, let us consider first the resummation procedure near the poles $\nu^2=4$ and $\nu^2=9$ and then generalize a result for an arbitrary pole $\nu^2=l^2,\ l\in \mathbb{Z}$.

\subsection{The pole structure near $\nu^2=4$} 
\noindent The superpotential has the following pole structure when $\nu^2\to 4$:
\begin{equation*}
\begin{split}
&\mathcal W(z)=\big[... \big]z+\bigg[\color{cornellred}\frac{\#}{\nu^2-4}\color{black}+\, ...\bigg]z^2+\bigg[\color{cadmiumgreen}\frac{\#}{\nu^2-4}\color{black}+\, ...\bigg]z^3+\bigg[\color{cornellred}\frac{\#}{(\nu^2-4)^3}\color{black}+\color{darkcerulean}\frac{\#}{\nu^2-4}\color{black}+\, ...\bigg]z^4+\\
&+\bigg[\color{cadmiumgreen}\frac{\#}{(\nu^2-4)^3}\color{black}+\, ...\bigg]z^5+\bigg[\color{cornellred}\frac{\#}{(\nu^2-4)^5}\color{black}+\color{darkcerulean}\frac{\#}{(\nu^2-4)^3}\color{black}+\, ...\bigg]z^6+...
\end{split}
\end{equation*} 
Different colours denote groups of terms, which are summed separately according to the resummation procedure suggested in \cite{GorskyMS}. Particularly, the sum of red terms, which are the most singular at each given order $\mathcal{W}_2,\,\mathcal{W}_4,\,\mathcal{W}_6$ and so on, is equal to 
\begin{equation*}
\displaystyle \bigg[g^{(2)}_{\, 1}\bigg(\frac{z}{2-\nu}\bigg)+g^{(2)}_{\, 1}\bigg(\frac{z}{2+\nu}\bigg)\bigg]z,
\end{equation*}
with the function $g^{(2)}_{\, 1}$ containing a logarithm:
\begin{equation*}
g_1^{(2)}(z)=\frac{1+\log(\frac{1}{2}(\sqrt{z^2+1}+1))-\sqrt{z^2+1}}{z}.
\end{equation*}
Sums of green and blue terms give functions $g^{(2)}_2$ and $g^{(2)}_{\, 3}$ correspondingly, which do not lead to branch cuts, and so on. So, we would like to know which intermediate states contribute to the red terms. As mentioned above, $\mathcal W_{N}$ is a sum of all possible terms $a_{\, k_1}^{n_1}a_{\, k_2}^{n_2}...\, a_{\, k_p}^{n_p}$ with $\sum k_i=-1, \sum n_i= N$. Degrees of leading singularities at $\nu^2=4$ in every coefficient $a^{n}_{\, k}$ are given in the following table (cf. (\ref{eq:coeff_exp_eps})):
\renewcommand*{\arraystretch}{1.2}
\begin{center}
  \begin{tabular}{ *{8}{ |c }}
     \multicolumn{1}{c}{} & \multicolumn{1}{c}{$\epsilon^{-4}$} & \multicolumn{1}{c}{$\epsilon^{-3}$} & \multicolumn{1}{c}{$\epsilon^{-2}$} & \multicolumn{1}{c}{$\epsilon^{-1}$} & \multicolumn{1}{c}{$\epsilon^0$} & \multicolumn{1}{c}{$\epsilon^1$} & \multicolumn{1}{c}{...}  \\ \hhline{*{4}{~}*{4}{|-}}
     \multicolumn{1}{c}{$H_1$} & \multicolumn{3}{ c |}{} & \cellcolor{cadmiumgreen!30} & \cellcolor{cadmiumgreen!30} & \cellcolor{cadmiumgreen!30} & \cellcolor{cadmiumgreen!30} \\ \hhline{*{3}{~}*{5}{|-}}
     \multicolumn{1}{c}{$H_2$} &\multicolumn{2}{ c |}{} & \cellcolor{cadmiumgreen!30} & $(\nu^2-4)^{-1}$ & $(\nu^2-4)^{-2}$ & $(\nu^2-4)^{-3}$ & \ \ ... \ \   \\ \hhline{*{2}{~}*{6}{|-}}
     \multicolumn{1}{c}{$H_3$} & \multicolumn{1}{ c |}{} & \cellcolor{cadmiumgreen!30} & 
     $(\nu^2-4)^{-1}$ & $(\nu^2-4)^{-2}$ & $(\nu^2-4)^{-3}$ & ...  \\ \hhline{*{1}{~}*{7}{|-}}
     \multicolumn{1}{c |}{$H_4$} & \cellcolor{cadmiumgreen!30} \ \ \ \ \ \ \ \ \ \ \  & $(\nu^2-4)^{-1}$ & $(\nu^2-4)^{-2}$ & $(\nu^2-4)^{-3}$ & ... &  \\ \cline{2-8}
     \multicolumn{1}{c |}{...} & & & & & &
  \end{tabular}
  \end{center}
  
\noindent More formally,
\begin{equation}
a_{k}^{n}\sim \frac{1}{(\nu^2-4)^{n+k}},\ \text{if}\ n\ge 2, \ n+k \ge 1\  (\text{uncoloured part of the table}),
\end{equation}
otherwise, if a coefficient $a_k^{n}$ has different labels from those listed above, it is in the coloured part of the table and it does not contain a pole at $\nu^2=4$. In total, the pole degree at $\nu^2=4$ of the product of coefficients $a_k^{n}$ with $\sum n_i=N,\ \sum k_i=-1$ is given by the following formula:
\begin{equation}
a_{\, k_1}^{n_1}a_{\, k_2}^{n_2}\dots a_{\, k_p}^{n_p}\sim (\nu^2-4)^{-(N-1)\ +\sum\limits_{\operatorname{coloured}}(n_i+k_i)}.
\end{equation}
The summation in the exponent is over all coefficients $a_k^{n}$ in the product, which are from the coloured part of the table. If $N$ is fixed, the presence of a multiplier $a^{n}_{\, k}$ from the coloured part of the table in the product increases the power of $(\nu^2-4)$ by $n+k\ge 0$ and makes the product less singular, except when $n=-k$. This means that the most singular power $(\nu^2-4)^{-(N-1)}$ in $\mathcal W_N$ is formed by those terms $a_{\, k_1}^{n_1}a_{\, k_2}^{n_2}...\, a_{\, k_p}^{n_p},\ \sum n_i =N$, which either do not include multipliers from the coloured part of the table or include only those of the form $a^{\ k}_{-k}$. It is worth mentioning, however, that poles of all degrees that are present in $\mathcal W_N$ get contributions from all the intermediate states $L_{-1}^k \ket{\Delta}$ with $k\le N$.

\subsection{The pole structure near $\nu^2=9$} 
\noindent Let us now consider the pole structure of the superpotential in the vicinity of the point $\nu^2=9$:
\begin{equation}
\label{eq:nu_3}
\begin{split}
&\mathcal W(z)=\big[... \big]z+\big[... \big]z^2+\bigg[\color{cornellred}\frac{\#}{\nu^2-9}\color{black}+\, ...\bigg]z^3+\bigg[\frac{\#}{\nu^2-9}+\, ...\bigg]z^4+\bigg[\frac{\#}{\nu^2-9}+\, ...\bigg]z^5+\\
&+\bigg[\color{cornellred}\frac{\#}{(\nu^2-9)^3}\color{black}+\, ...\bigg]z^6+...
\end{split}
\end{equation} 
As above, the red terms, which are the most singular in the superpotential coefficients $\mathcal{W}_3,\,\mathcal{W}_6,\mathcal{W}_9$ and so on, sum up into the function 
\begin{equation*}
\displaystyle \bigg[g^{(3)}_{\, 1}\bigg(\frac{z^{3/2}}{3-\nu}\bigg)+g^{(3)}_{\, 1}\bigg(\frac{z^{3/2}}{3+\nu}\bigg)\bigg]z^{3/2},
\end{equation*}
containing a logarithm. However, leading singularities in every coefficient $a^{n}_{\, k}$ when $\nu^2 \to 9$ are now described by a slightly different table:
  \renewcommand*{\arraystretch}{1.2}
  \begin{center}
  \begin{tabular}{ *{9}{| c }}
     \multicolumn{1}{c}{} & \multicolumn{1}{c}{$\epsilon^{-5}$} & \multicolumn{1}{c}{$\epsilon^{-4}$} & \multicolumn{1}{c}{$\epsilon^{-3}$} & \multicolumn{1}{c}{$\epsilon^{-2}$} & \multicolumn{1}{c}{$\epsilon^{-1}$} & \multicolumn{1}{c}{$\epsilon^0$}  & \multicolumn{1}{c}{$\epsilon^1$}& \multicolumn{1}{c}{...}  \\ \hhline{*{5}{~}*{4}{|-}}
     \multicolumn{1}{c}{$H_1$} & \multicolumn{4}{ c |}{} & \cellcolor{cadmiumgreen!30} & \cellcolor{cadmiumgreen!30} & \cellcolor{cadmiumgreen!30} & \cellcolor{cadmiumgreen!30}   \\ \hhline{*{4}{~}*{5}{|-}}
     \multicolumn{1}{c}{$H_2$} &\multicolumn{3}{ c |}{} & \cellcolor{cadmiumgreen!30} & \cellcolor{cadmiumgreen!30} & \cellcolor{cadmiumgreen!30}  & \cellcolor{cadmiumgreen!30}  & \cellcolor{cadmiumgreen!30}   \\ \hhline{*{3}{~}*{6}{|-}}
     \multicolumn{1}{c}{$H_3$} & \multicolumn{2}{ c |}{} & \cellcolor{cadmiumgreen!30} & \cellcolor{cadmiumgreen!30} & $(\nu^2-9)^{-1}$ & $(\nu^2-9)^{-2}$ & $(\nu^2-9)^{-3}$ & \ \ ... \ \   \\ \hhline{*{2}{~}*{7}{|-}}
     \multicolumn{1}{c }{$H_4$} & \multicolumn{1}{ c |}{} & \cellcolor{cadmiumgreen!30} & \cellcolor{cadmiumgreen!30} & $(\nu^2-9)^{-1}$ & $(\nu^2-9)^{-2}$ & $(\nu^2-9)^{-3}$ & \ \ ... \ \ &   \\ \hhline{*{1}{~}*{8}{|-}}
       \multicolumn{1}{c |}{$H_5$} & \cellcolor{cadmiumgreen!30} \ \ \ \ \ \ \  \ \ \ \ & \cellcolor{cadmiumgreen!30} \ \ \ \ \ \ \ \ \ \ \ & $(\nu^2-9)^{-1}$ & $(\nu^2-9)^{-2}$ & $(\nu^2-9)^{-3}$ & \ \ ... \ \ & &   \\ \cline{2-9}
     \multicolumn{1}{c |}{...} & & & & & & &
  \end{tabular}
  \end{center}
  
\noindent Similar to the previous case,
\begin{equation}
a_{k}^{n}\sim \frac{1}{(\nu^2-9)^{n+k-1}},\ \text{if}\ n\ge 3, \ n+k \ge 2\ 
(\text{uncoloured part of the table}),
\end{equation}
otherwise, if a coefficient $a_k^{n}$ has different labels from those listed above, it belongs to the coloured part of the table and it does not have a pole at $\nu^2=9$. Thus, the pole degree at $\nu^2=9$ of the product of coefficients $a_k^{n}$ with $\sum n_i=N,\ \sum k_i=-1$ is given by the following formula:
\begin{equation}
\begin{split}
&a_{\, k_1}^{n_1}a_{\, k_2}^{n_2}\dots a_{\, k_p}^{n_p}\sim (\nu^2-9)^{-(N-2)\, +\, \big[\# (\operatorname{uncoloured})-1\ +\sum\limits_{\operatorname{coloured}}(k_i+n_i)\big]}.
\end{split}
\end{equation}
The summation in the exponent is over all coefficients $a_k^{n}$ in the product, which are from the coloured part of the table, and $\#(\operatorname{uncoloured})$ is a number of coefficients $a_k^n$ from the uncoloured part of the table in the product. As in the previous case, every multiplier from the coloured part of the table increases the power of $(\nu^2-9)$ by $n+k\ge0$. The difference from the previous case is that the more coefficients $a^{n}_{\, k}$ from the uncoloured part of the table present in the product the less singular it is. So, the products of coefficients $a^{n}_{\, k}$, satisfying the following conditions, contribute to the most singular terms in $\mathcal W_{3},\, \mathcal W_{6},\, \mathcal W_{9}$ and so on:
\begin{equation*}
\begin{split}
&\mathcal W_3:\ \# (\operatorname{uncoloured})-1\ +\sum\limits_{\operatorname{coloured}}(k_i+n_i)=0 \\
&\mathcal W_6:\ \# (\operatorname{uncoloured})-1\ +\sum\limits_{\operatorname{coloured}}(k_i+n_i)=0,1 \\
&\mathcal W_9:\ \# (\operatorname{uncoloured})-1\ +\sum\limits_{\operatorname{coloured}}(k_i+n_i)=0,1,2
\end{split}
\end{equation*}
Again, all the intermediate states $L_{-1}^k \ket{\Delta},\ k\le N$ contribute to the superpotential coefficient $\mathcal{W}_N$.

\subsection{The pole structure near $\nu^2=l^2,\ l\in \mathbb{Z}$}
\noindent The results above can be easily generalized. For the pole $\nu^2=l^2$ the leading singularities from the superpotential coefficients $\mathcal W_l, \mathcal W_{2l}, \mathcal W_{3l},...$ sum up into the logarithm. In this case the leading behaviour of the coefficients $a_k^{n}$ is as follows:
\begin{equation}
a_{k}^{n}\sim \frac{1}{(\nu^2-l^2)^{n+k-l+2}},\ \text{if}\ n\ge l, \ n+k \ge l-1,
\end{equation}
and if a coefficient $a_k^{n}$ has labels, not satisfying conditions above, then it does not have a pole at $\nu^2=l^2$. Similar to the previous cases, we call a coefficient $a_k^n$ uncoloured if its indices $(n,k)$ satisfy the condition $n\ge l, \ n+k \ge l-1$ and coloured otherwise. Thus, the pole degree at $\nu^2=l^2,\ l\in \mathbb{Z}$ of the product of coefficients $a_k^{n}$ with $\sum n_i=N,\ \sum k_i=-1$ is given by the following formula
\begin{equation*}
\begin{split}
&a_{\, k_1}^{n_1}a_{\, k_2}^{n_2}\dots a_{\, k_p}^{n_p}\sim (\nu^2-l^2)^{-(N-l+1)\, +\, \big\{(l-2)\big[\# (\operatorname{uncoloured})-1\big]\ +\sum\limits_{\operatorname{coloured}}(k_i+n_i)\big\}}.
\end{split}
\end{equation*}
As in the previous cases, the most singular pole in $\mathcal W_N$ is composed of the terms $a_{\, k_1}^{n_1}a_{\, k_2}^{n_2}...\, a_{\, k_p}^{n_p}$, which contain a minimal amount of multipliers from the uncoloured part of the table and a "minimal"\  contribution of multipliers from the coloured one. The products of coefficients $a^{n}_{\, k}$ contributing to the most singular term in $\mathcal W_{n\cdot l},\, n\in \mathbb N$ satisfy the following condition:
\begin{equation*}
(l-2)\big[\# (\operatorname{uncoloured})-1\big]\ +\sum\limits_{\operatorname{coloured}}(k_i+n_i)=0, \dotsc,(n-1)(l-2).
\end{equation*}
In general, poles of all degrees that are present in $\mathcal W_N$ get contributions from all the intermediate states $L_{-1}^k \ket{\Delta}$ with $k\le N$.

\section{Conclusion}
In this paper we studied the irregular conformal block and identified contributions of descendants of the intermediate state to the most singular terms in each instanton sector of the superpotential, which are responsible for the appearance of the branch cuts after resummation. It was known \cite{MMM} that only descendants of the form $L_{-1}^N\ket{\Delta}$ contribute to the irregular conformal block, if it is considered as the limit of the 4-point conformal block. Thus, we showed which of them are responsible for the branch cuts in the superpotential, namely the most singular pole contained in the superpotential expansion coefficient $\mathcal W_N$ gets contributions from all the intermediate states $L_{-1}^k \ket{\Delta}$ with $k\le N$. 

Two possibilities exist to pose a question about resummation of the superpotential by translating it into the language of AdS${}_3$/CFT${}_2$ correspondence. First, one may try to interpret the resummation procedure in terms of resummation of global blocks, or, equivalently, of geodesic Witten diagrams. It was established in the work of Hijano and coauthors \cite{Hijglob} that a geodesic Witten diagram, in which the integration over positions of trivalent vertices in the bulk goes along geodesics connecting boundary points, corresponds to the global conformal block, containing contributions only from the global conformal family of the intermediate state. It is known also that the Virasoro conformal block can be represented as a sum of global conformal blocks, which are just hypergeometric functions \cite{BPZ} (Appendix B).

Alternatively, the conformal block in the heavy-light limit can be interpreted as the geodesic motion in AdS${}_3$ in the background created by the heavy boundary insertions and this limit is well-studied \cite{WSBH}. This interpretation becomes completely clear when one utilizes the monodromy method to evaluate the conformal block. The classical conformal block in this picture coincides with the action of a particle moving in the black hole geometry, created by heavy operators. However, the limit of the conformal block one have to deal with to understand the resummation procedure on the side of $AdS_3$ is different from the heavy-light limit, namely $\Delta_i\sim c,\ c\to \infty$ and $\Delta_{2,3}/c \to \infty$ in the case considered in the paper. Thus, the first problem one has to solve to get an interpretation of the resummation procedure in AdS${}_3$ is to find an object in AdS${}_3$ gravity, dual to the irregular conformal block. It is known that the Mathieu equation is an analogue of the monodromy equation for the irregular conformal block \cite{Jap,Piatek}. So, we could suggest that in order to obtain the interpretation of the Gaiotto state or of the irregular conformal block, one needs to focus on the Mathieu equation.

We are grateful to A. Gorsky for suggesting this problem and for numerous fruitful discussions. The work of S.A and M.L was supported by Basis Foundation Fellowship.

\bibliographystyle{unsrt}
\bibliography{references}

\end{document}